# Large Area Metalenses:
# Design, Characterization, and Mass Manufacturing

**Alan She, Shuyan Zhang, Samuel Shian,
David R. Clarke, and Federico Capasso[*]**

*John A. Paulson School of Engineering and Applied Sciences, Harvard University,
Cambridge, MA 02138*
*[*]capasso@seas.harvard.edu*

**Abstract:** Optical components, such as lenses, have traditionally been made in the bulk form by shaping glass or other transparent materials. Recent advances in metasurfaces provide a new basis for recasting optical components into thin, planar elements, having similar or better performance using arrays of subwavelength-spaced optical phase-shifters. The technology required to mass produce them dates back to the mid-1990s, when the feature sizes of semiconductor manufacturing became considerably denser than the wavelength of light, advancing in stride with Moore's law. This provides the possibility of unifying two industries: semiconductor manufacturing and lens-making, whereby the same technology used to make computer chips is used to make optical components, such as lenses, based on metasurfaces. Using a scalable metasurface layout compression algorithm that exponentially reduces design file sizes (by 3 orders of magnitude for a centimeter diameter lens) and stepper photolithography, we show the design and fabrication of metasurface lenses (metalenses) with extremely large areas, up to centimeters in diameter and beyond. Using a single two-centimeter diameter near-infrared metalens less than a micron thick fabricated in this way, we experimentally implement the ideal thin lens equation, while demonstrating high-quality imaging and diffraction-limited focusing.

## 1. Introduction

Metasurfaces control the wavefront of light using arrays of fixed optical phase shifters, amplitude modulators, and/or polarization changing elements[1]. These are patterned on a surface to introduce a desired spatial distribution of optical phase, amplitude, and/or polarization. By tailoring the properties of each element of the array, one can spatially control these properties of the transmitted, reflected, or scattered light and consequently mold the wavefront[2]. Based on this concept, various functionalities have been demonstrated including lenses, axicons, blazed gratings, vortex plates and wave plates[3–11]. These devices are thin and lightweight. In order to maximize performance, metasurfaces utilize arrays of elements with subwavelength pitch, such that the optical wavefront is unaffected by the discretization of the elements. Notably, multiple optical functions can be encoded in a metasurface phase profile[12]. Metasurfaces are phased-array antennas taken to the extreme - at shorter wavelengths and with much smaller feature sizes. The planarity of metasurfaces allows for fabrication routes directly in line with conventional processes of the mature integrated circuit (IC) industry, allowing for opportunities to a scale unmet by 3D designs.

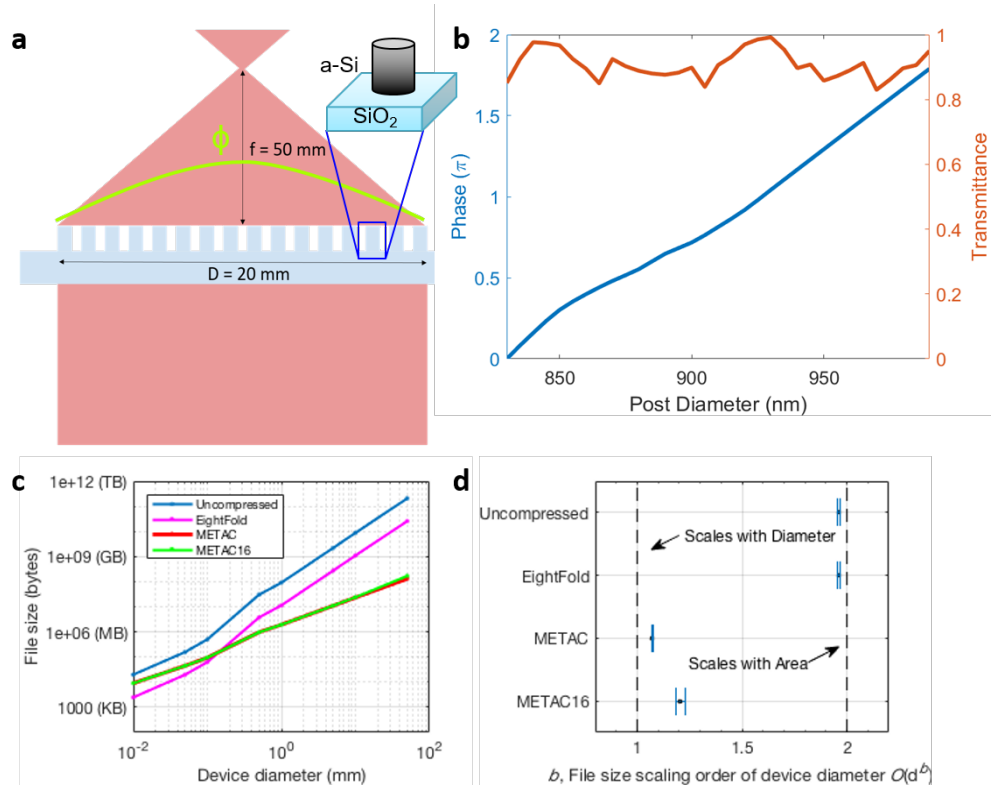

**Fig. 1. Metasurface lens design. (a) A schematic shows a metasurface lens (metalens) that is designed to focus light of normal incidence, where D is diameter and f is focal length. The phase profile, φ, is implemented by a dense array of microscopic meta-elements, made of amorphous silicon cylindrical posts, supported by a SiO$_2$ substrate. (b) A range of diameters of meta-element is used and their corresponding phase response (blue) and transmittance (red) are plotted in the graph. (c) Due to the extremely large number of meta-elements required to comprise a large area metalens, a metasurface data compression algorithm, which we call METAC, was developed to generate manageable file sizes of metalens designs. Four methods were compared: uncompressed, EightFold (design is divided into copies of eighths), METAC, and METAC16 (maximum number of levels is restricted to 16, for better compatibility with existing software). File size is plotted as a function of device diameter. (d) The scaling order, b, of the file size for each method is plotted. Error bars represents one standard deviation. Methods with b closer to 2 indicate file size scaling with device area, while those with b closer to 1 indicates scaling with device diameter.**

## 2. Lens design

To demonstrate large area metasurface optics using a process that could scale to high volumes and low costs, we designed and fabricated a transmissive metasurface lens (metalens) 2 cm in diameter. In order to produce this lensing effect, the metalens imposed a spatial profile of phase shifts (phase profile)[13] on the wavefront is $\varphi = -2\pi/\lambda[(r^2 + f^2)^{1/2} - f]$, where λ=1550 nm is the design wavelength (1550 nm), *r* is the radial coordinate, and *f=50* mm is the focal length (Fig. 1a). The wavelength was chosen based on feasibility of fabrication within the limitations of the equipment available in our labs. Amorphous silicon (a-Si) on silica (SiO$_2$) was chosen as the metalens material. This phase profile focuses collimated light into a spot and was constructed using a dense pattern of meta-elements, each of which acted as a miniature antenna to locally impart a controlled phase shift. We used cylindrical posts, given their polarization-independent response, with diameters ranging from 830-990 nm and a fixed height of 600 nm (Fig. 1b). By varying the diameter, the posts were able to produce 2π phase coverage and high, relatively uniform transmittances. To maximize optical efficiency, the placement of meta-elements was made denser using fixed edge-to-edge separation instead of the traditional fixed center-to-center separation[14]. In our design, we chose the edge-to-edge spacing to be 650 nm considering the fabrication limitations and avoiding the interaction between neighboring antennas. The numerical simulation was carried out using the finite difference time domain method (FDTD module, Lumerical Inc.) (Fig. A1). Because of the memory space limitation, we simulated a smaller lens with the same NA as the fabricated (ø=100 μm, NA = 0.07).

With metasurfaces, the data describing large designs are faced with the challenge of enormous file sizes due to having millions or billions of individual microscopic meta-elements (necessitated by the subwavelength size criterion) described over macroscopically large device areas. This extremely high data density over large areas generates unmanageably large total file sizes, limiting the fabrication of metalenses to sizes no larger than a few millimeters. For example, a 5 cm diameter device may be comprised of over 6 billion meta-elements, each instance of which must be described by nanometer-precision definitions of position and radius, resulting in a size >200 gigabytes (Table 1). Since these design files must subsequently undergo computationally intensive processes such as data conversion (often known as "fracturing") for use with mask writing equipment, it is critical that these file sizes be minimized. We have implemented a compression algorithm that generates these design files, which we refer to as "METAC" (METAsurface Compression), allow the file size to be reduced by many orders of magnitude (Fig. 1c). The algorithm uses a large number of hierarchical levels in the layout file (e.g., GDSII) to copy meta-elements about a central optical axis (most lenses are centrosymmetric), leveraging rotational symmetry to greatly reduce the number of unique meta-element definitions to those along a single radial line. In some cases, certain conversion software or mask writing machines impose an upper limit on the number of levels that any design may contain (e.g., 16, which was used with our design), so an algorithm based on METAC but limited to 16 levels, which we call METAC16, was included for reference.

**Table 1. Design file size according to device diameter and comparison of METAC algorithm**

| Device diameter | Meta-element count | Uncompressed file size (B=Byte) | METAC file size |
|---|---|---|---|
| 10 μm | 150 | 19.8 kB | 9.1 kB |
| 50 μm | 3,614 | 157.9 kB | 44.5 kB |
| 100 μm | 14,068 | 501.4 kB | 94.1 kB |
| 500 μm | 1,053,822 | 30.3 MB | 972.1 kB |
| 1 mm | 3,204,089 | 91.2 MB | 1.9 MB |
| 5 mm | 73,194,422 | 2.2 GB | 11.0 MB |
| 10 mm | 291,697,949 | 8.8 GB | 23.3 MB |
| 50 mm | 6,853,721,364 | 205.7 GB | 131.1 MB |

We analyzed the METAC algorithm in comparison to other methods of data representation (Fig. 1c,b and Table 1), including uncompressed, EightFold, and METAC16. Uncompressed is the method where every meta-element is defined explicitly. Eightfold is the method where the entire design is comprised of eight copies of a single unique octant so that its file size is expectedly one-eighth of the uncompressed. We see that METAC and METAC16 have comparable performance and are extremely effective at compressing file sizes as device diameter increases to the millimeter regime and beyond. We compare a file size scaling order, $b = \log_d F$, which determines the rate at which the file size increases with device dimensions, where $F$ is file size and $d$ is the device diameter (Fig. 1d). The uncompressed file size scales with the square of the diameter ($b \approx 2$), simply because of the element count scales with device area, and similarly with EightFold. Because of this quadratic scaling, the file size of metasurface designs quickly reach the gigabyte or even terabyte regime for centimeter scale devices. Ideally, with a metasurface lens, which is highly centrosymmetric, unique data is contained only along the radial dimension, i.e., ideally linear or $b = 1$. In fact, METAC yields $b = 1.073$, with 0.073 attributed to a data definition and reference overhead. Even with METAC16, $b$ is increased slightly to 1.207, since extra data is required to compensate for the level limit. In the analysis here, the average Voronoi cell length of meta-elements was 520 μm, which is on the large side, indicating that performance gains will be even more pronounced for denser settings. We believe that METAC algorithm will facilitate the development of large area metasurfaces by preventing debilitatingly large file sizes in the realm of terabytes and petabytes as feature densities increase and device areas continue to grow towards the centimeter scale and beyond.

## 3. Fabrication methodology

Using a design generated by the METAC algorithm, we fabricated metalenses on a 4-inch fused silica (SiO$_2$) wafer substrate (Fig. 2). A film stack was created over the cleaned substrate, comprised of (from bottom to top) 0.6 μm of a-Si, 1.1 μm of SPR700-1.0 (MEGAPOSIT), and 0.4 μm of CEM365IS (ShinEtsuMicroSi). The a-Si layer was deposited using plasma-enhanced chemical vapor deposition (PECVD by STS). The SPR700-1.0 and CEM365IS were spun-coated at 4000 rpm. The wafer was then exposed using a GCA AS200 i-line stepper. The exposure was followed by a water rinse and post-exposure bake at 115 °C for 60 seconds. The wafer was then developed in MF CD-26 (Shipley) for 60 seconds in two baths and immediately rinsed in water. The pattern was then etched into the a-Si using reactive ion etching (STS MPX/LPX ICP RIE). Finally, the resist was stripped by soaking in Remover PG for 10 hours, followed by 2 minutes using the Matrix 105 Plasma Asher (1500 mTorr oxygen pressure, 500 W RF power, 200 °C).

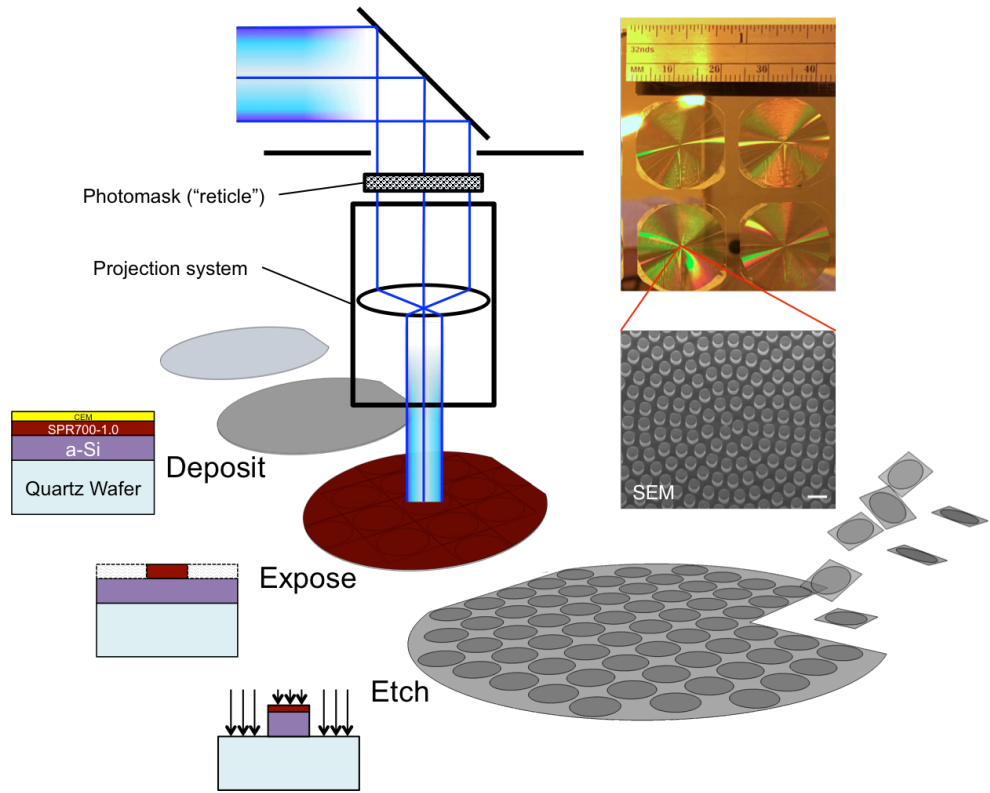

**Fig. 2.** A schematic diagram shows the production of metalenses at low cost and with high yield using existing photolithographic stepper technology. Here a wafer substrate is first deposited with the appropriate film stack, comprised of the metalens material (amorphous (a)-Si), photoresist (SPR700-1.0), and contrast enhancement material (CEM). The pattern of the metalens, which is contained in the reticle, is then projected by the stepper, and replicated rapidly over the face of the wafer by repeatedly exposing and incrementally stepping the wafer position. Throughputs as high as hundreds of wafers per hour (wph) can be achieved. Then the pattern is etched into the a-Si, forming the metalens. Finally, after any residual photoresist is removed, the wafer can be diced into separate individual metalens devices. A photo of fabricated metalens (upper right), 2-cm in diameter, using this methodology is shown in comparison to a ruler. A SEM of the metalens center (center right) shows the microscopic posts comprising the metalens. Scale bar: 2 μm.

### 4. Flatness requirement of large area metasurface lenses

For large area devices, the question of how substrate flatness affects optical performance becomes relevant. Since metasurfaces are usually referred to as phase engineered devices, it is often assumed that strict tolerances must be imposed on phase, and therefore substrate flatness variations must be kept much less than the wavelength over the entire span of the device. However, we have found that tolerances are application dependent, and in particular applications, such as lensing, these tolerances can be relaxed, as opposed to other applications such as holograms and interferograms.

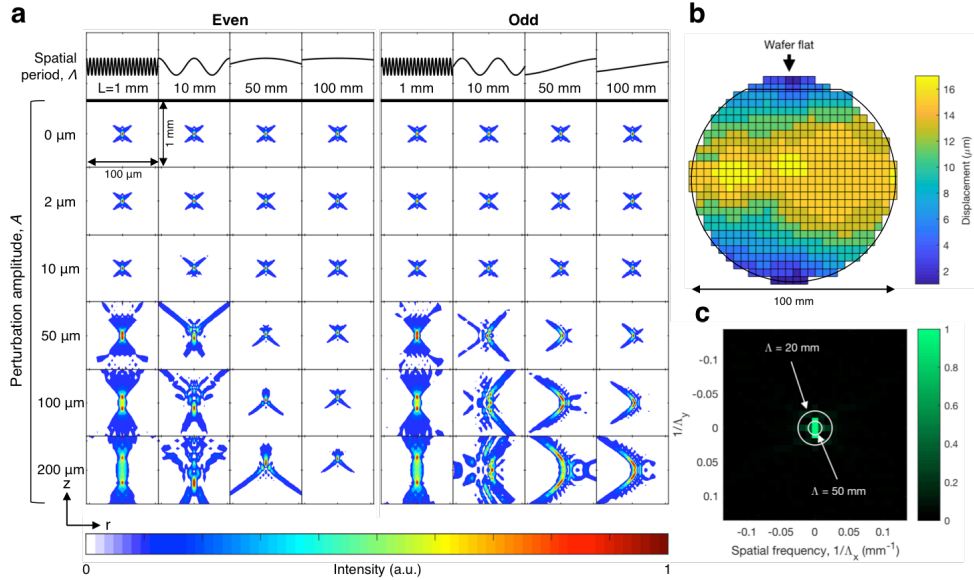

**Fig. 3. Flatness requirements. (a) As metalens size increases, the flatness over the device surface becomes increasingly relevant. The flatness requirements were studied by computer simulations of metalenses (diameter: 2 cm, focal length: 50 mm, design wavelength: 1550 nm), in which the surface curvature was varied (spatial period ranging from 1 to 100 mm, and perturbation amplitude from 0 to 200 μm). The intensity of the optical field is shown. The metalens is situated at the bottom of each plot, and the vertical (z) and horizontal (r) axes are the propagation direction and radial dimension, respectively. Each group four columns on the left and right show the optical behavior for even (cosine) and odd (sine) spatial perturbations, respectively, with respect to the metalens center. Quality of focus is reduced for shorter spatial periods and higher perturbation amplitudes. In (b), the surface profile of the 4-inch wafer we used (including metalenses) was measured (using Toho FLX-2320-S) and the Fourier transform calculated in (c) to obtain the major contributions to spatial frequencies, which mainly occurred at $\Lambda^{-1} < 0.02$ mm$^{-1}$ (or $\Lambda > 20$ mm). The inner and outer white circles denote $\Lambda$ at 50 and 20 mm, respectively.**

Substrate flatness can be characterized in terms of a surface profile, which are commonly met in various contexts as curvature (warp and bow), total thickness variation, and surface roughness. These properties may be quantified, which exist within a spectrum surface displacement profiles, $\Delta z(x, y)$, with two parameters: spatial period (or frequency), $\Lambda$, and perturbation amplitude, $A$, (Fig. 3) where surface roughness, warp/bow, and total thickness variation occur with increasing spatial periodicity. Mathematically, we can consider even and odd components, where $\Delta z = A\cos(2\pi x/\Lambda)$ or $A\sin(2\pi x/\Lambda)$, respectively, and generally any profile can be represented as a Fourier series of these. The effect of the displacement profile can be understood in terms of both ray and wave optics. In view of ray optics, metasurfaces bend light at the device surface, such that the outgoing angle is determined by the angle of incidence and phase gradient at the position of interaction. In wave optics, the displacement profile given coherent illumination modifies interference and is less of an issue when using incoherent light. By computer simulations of a representative large area metalens (diameter: 2 cm, focal length: 50 mm, design wavelength: 1550 nm), we found that the quality of focus worsened as $\Lambda$ decreased and $A$ increased, due to more pronounced angular deviations and interference (Fig. 3a). As a rule of thumb, good quality of focus (i.e., the substrate is "adequately flat") is seen for A ≤ 10 μm and $\Lambda$ > 10 mm (Fig. 3a). We measured the surface profile of our wafer as a representative sample, following processing of an array of metalenses, to be $A$ < 17 μm (Fig. 3b) and $\Lambda$ > 20 mm (Fig. 3c), satisfying the adequate flatness criteria given by the simulations.

## 5. Experiment

The focal spot and imaging performance of the metalens were characterized using a horizontal microscope setup (Fig. A2) and simple imaging setup (Fig. 4a). A tunable laser (HP 8168F) operating between λ=1440-1590 nm with an optical fiber collimator (Thorlabs F810APC-1550) produced a beam 7 mm in diameter, illuminating the metalens center (while the numerical aperture (NA) of the metalens is 0.2, the limited size of the illuminated area results in an effective NA=0.07). The focal spot was imaged by a horizontal microscope: an objective (10x Mitutoyo M Plan Apo NIR infinity corrected objective), tube lens (Plano-Convex Lens, f = 200.0 mm), and camera (digital InGaAs, Raptor OWL640). The entire horizontal microscope was mounted on a linear motor (NPM Acculine SLP35), which allowed horizontal scanning of the light field. At λ=1550 nm, the focal length of the device was measured to be 50.159 ± 0.023 mm (design focal length = 50 mm), and the full-width ($1/e^2$) of the focal spot was measured to be 20.9 μm (Fig. 4b). The theoretical diffraction-limited spot size of this device assuming an aperture of 7 mm and perfect gaussian illumination ($M^2$ = 1) is 14.1 μm. The modulation transfer function (MTF) was calculated by the Fourier transform of the point spread function (i.e., focal spot image) and showed good agreement with the theoretical diffraction-limited MTF (Fig. 4c).

The focusing efficiency was determined by the ratio of total optical power with and without the metalens measured at the focus position with an optical power meter (Thorlabs PM100D). A pair of irises were used to block any stray light. The efficiency was measured to be 91.8 ± 4.1% at λ=1550 nm, which does not include losses due to the air-silica boundary (3.25% loss by normal incidence Fresnel reflection). The efficiency may be further improved (e.g. to > 95%) by adding an antireflection coating.

The effect of chromatic aberration was studied in two ways: chromatic focal shift and imaging chromatic aberration, by sweeping the source wavelength while varying and fixing the camera position, respectively. The chromatic focal shift was highly linear ($R^2$=0.9982) and measured to be $\Delta f/\Delta\lambda$= -0.0335 mm/nm, corresponding to a 10% focal shift (-5.09 mm) as the wavelength was tuned by 150 nm (1440-1590 nm) (Fig. 4d). A hyperspectral image was taken of the focal spot (Fig. 4e) to see the imaging chromatic aberration. The focal spot appears white with a faint tinge of color at the periphery, indicating a minor amount of chromatic aberration, which is attributed to the low NA. This is corroborated by the horizontal and vertical line cuts, which coincide closely.

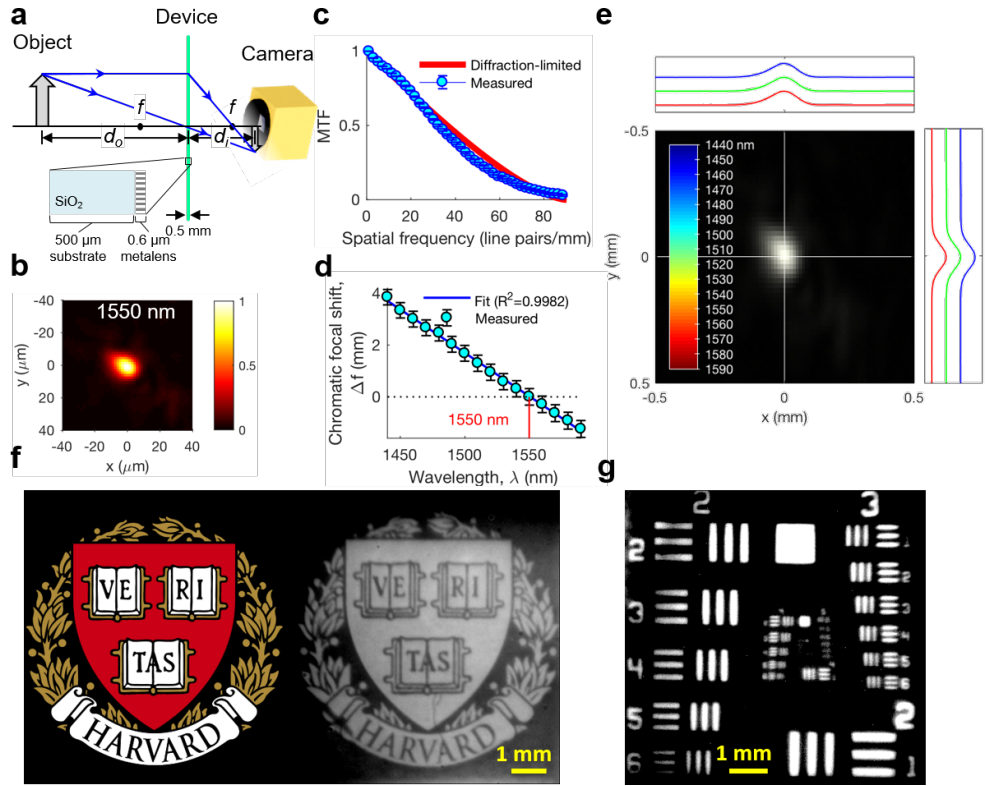

**Fig. 4. Focusing and imaging performance.** (a) The thinness of the device allows for imaging setups very similarly to the ideal thin lens equation, which was used to demonstrate imaging capabilities. (b) Image of focal spot with 7 mm gaussian illumination at λ=1550 nm. (c) The measured modulation transfer function (MTF) from (b) is plotted with the theoretical diffraction-limited MTF. Error bars: standard deviation. (d) Chromatic focal shift was measured as a function of the wavelength of illumination. The measured deviation of focal length from that of the design wavelength at 1550 nm (light blue dots, error bars: standard deviation) is plotted with the linear fit (blue line). (e) Hyperspectral image of focal spot in the same configuration as (b) for λ=1440-1590 nm in 10 nm increments linearly binned to RGB channels (center wavelengths $\lambda_R$=1590, $\lambda_G$=1515, and $\lambda_B$=1480 nm). The spot, which is largely white, indicates little chromatic aberration, which can be attributed to the low NA (0.07). Horizontal and vertical line cuts at the RGB center wavelengths are also shown. Using the thin lens setup in (a), simple, single-lens imaging was demonstrated at λ=1550 nm for (f) the Harvard university logo and (g) US Air Force 1951 resolution target, without any additional optical components.

The above setup was modified to perform direct single lens imaging of macroscopic objects (Fig. 4a). Due to the thinness of the metalens, this configuration closely approximated that of the traditional thin lens equation[15], which is commonly taught in introductory optics textbooks yet does not accurately describe the imaging conditions of real, bulky lenses, since it assumes a lens of negligible thickness. The entire setup consisted of an illuminated object, the metalens, and camera. Objects were patterns made using chromium on glass, in the form of the Harvard University logo and USAF 1951 resolution target (Fig. 4f,g respectively). The images were brought into focus by adjusting the object-to-lens distance ($d_o$) to 175.0 ± 0.5 mm, and lens-to-camera distance ($d_i$) to 70.0 ± 0.5 mm. This agrees with the thin lens equation ($1/d_i + 1/d_o = 1/f$, where and *f* is the focal length) for a lens with $f$ = 50 mm. The optical magnification was measured to be -0.42, since a 4 mm reference object was inverted and imaged to have a length of 1.69 mm on the camera sensor. The corresponding calculated value according to the thin lens magnification formula, $M = f/(f - d_o) = -0.40 \pm 0.003$.

## 6. Conclusion

We have shown the design, fabrication, and optical characterization of the large area metalenses, 2 centimeters in diameter with efficiency greater than 91%, fabricated using photolithographic steppers. Our algorithm for large area, high data density metasurface designs is general: in addition to the photolithographic method described, the algorithm can be used for other fabrication methods, such as nanoimprint and self-assembly based techniques[16–18], where highly symmetric patterns are utilized. The metalens, which is only 600 nm thick, in combination with the 0.5 mm thick substrate is a close experimental approximation of a lens described by the ideal thin lens equation. Optical components produced using this method have several advantages. In contrast to machining, such as diamond turning[19] and magnetorheological finishing[20], which form the shape of lenses very precisely but relatively slowly, photolithography can imprint properties of an optical component with a flash exposure only milliseconds in duration[21]. Also, in contrast to molds[22], a projection photomask can hypothetically be used an unlimited number of times without any wear or loss of fidelity. Finally, fabrication processes are shouldered by mature semiconductor manufacturing technology. Moore's law, which describes the trend where the number of transistors per square inch doubles every year[23], is mainly defined, among other factors, by the rate at which lithographic techniques can shrink down pattern feature sizes, i.e. technology nodes sizes[24–26] (Fig. A3). In 1992, the technology node size (TNS) dropped below the 700 nm, which is coincidentally the red end of the visible spectrum[27]. In 1995, TNS achieved 350 nm, providing twice the feature density and sufficient for subwavelength sampling of 700 nm wavelength light – from that point forward, the mass production of subwavelength metasurfaces working at visible wavelengths became possible. Using high-yield technologies such as photolithographic steppers and scanners[28], we envision a manufacturing transition from using machined or molded optics to lithographically patterned optics, where they can be mass produced with the similar scale and precision as IC chips in semiconductor fabrication facilities, such as foundries. An arrangement, in which optical and electronic parts are produced in the same place with the same equipment, is also naturally advantageous for integration. State-of-the-art equipment is useful, but not necessarily required. In particular, as it becomes ever more difficult to keep pace with Moore's law at the time of writing and as state-of-the-art lithography systems, such as extreme ultraviolet lithography (EUVL)[29], become so overwhelmingly expensive such that they are inaccessible to everyone except for the largest players in the industry, it is important to find avenues in which existing capital equipment can be brought out of obsolescence and repurposed for new, exciting opportunities.

# 7. Appendix

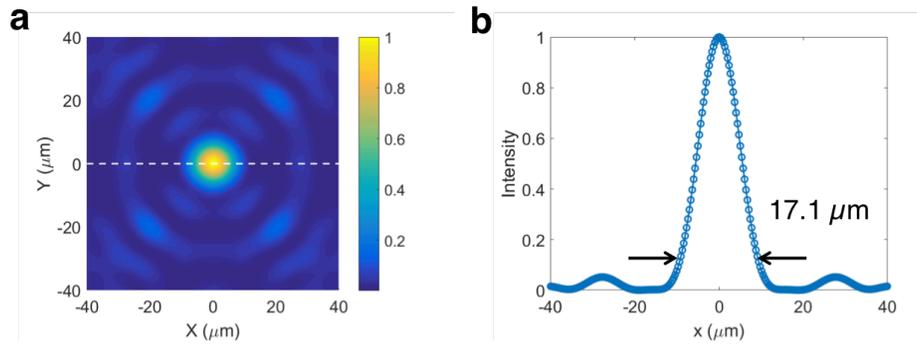

**Fig. A1.** FDTD simulations of focal spot of a 2 cm diameter metalens: (a) Simulated distribution of the electric field intensity (normalized $|E|^2$) of the focal spot. (b) The cross section of intensity profile at Y = 0 (white dashed line in (a)) from which the size of the focused beam can be determined. This simulation indicates that the focal spot size, i.e. the full beam waist at $1/e^2$ of the peak intensity, is 17.1 µm.

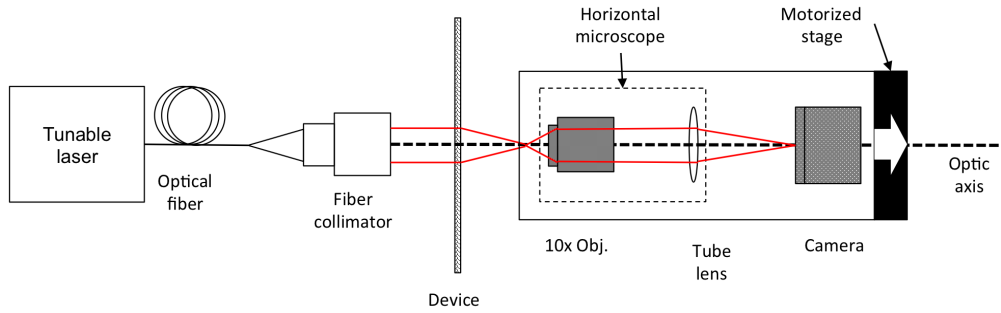

**Fig. A2.** A schematic diagram shows the setup for characterization of focal spot. A horizontal microscope and a camera are mounted on a motorized stage to scan the image the focal spot, which later can be 3D constructed to determine the focal length and the spot size of a metalens device.

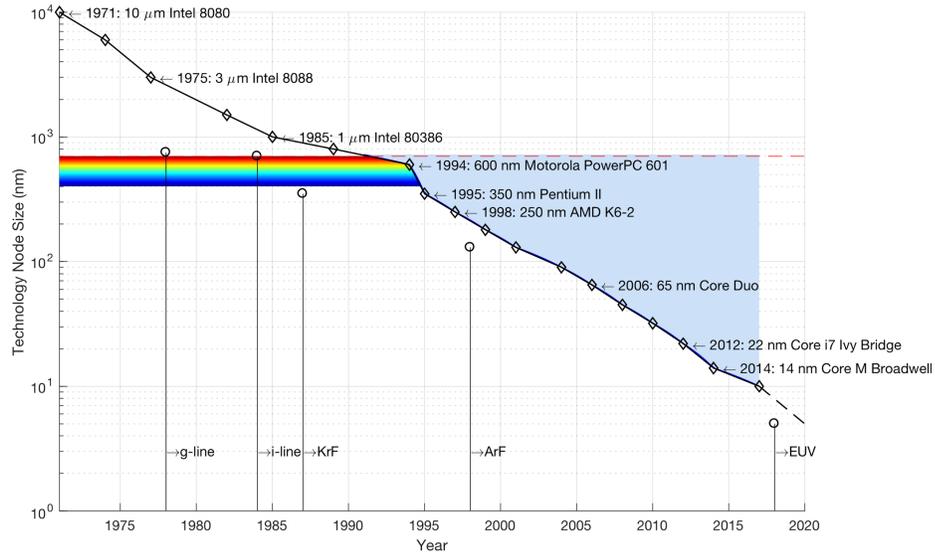

**Fig. A3.** Being an early enabler of metasurfaces, Moore's law, which predicts the transistor areal density in computer chips to double every year, is driven in large by improvements in lithographic technology. The plot shows the state of lithographic technology, represented as technology node size (TNS, diamond symbols), as a function of year, where smaller TNSs indicate higher feature densities. Several important product landmarks utilizing these TNSs are labelled, and key leaps in TNS powered by new light sources are denoted by vertical stems with circles. These past developments will enable the possibility of large area, cost-efficient metasurfaces optical devices. The red dotted line denotes the 700 nm threshold (corresponding to red light), below which the TNS had already surpassed in the mid-1990s. The wavelengths of the visible spectrum corresponding to TNS are shown shaded from 400-700 nm. In general, subwavelength-sized features can be produced using TNSs with at least twice the feature density as compared to the wavelength of light. Subsequent improvements to TNS up to the present (blue shaded region) have enabled feature sizes much smaller than the wavelength of light, providing ways to create more complex, fine-featured meta-elements as well as alternative routes for effectively utilizing foundry equipment which would otherwise be viewed as obsolete.

**Funding**

Air Force Office of Scientific Research (AFOSR) (MURI: FA9550-12-1-0389); Charles Stark Draper Laboratory Fellowship; A*STAR Singapore National Science Scholarship; National Science Foundation (NSF) (CMMI-1333835, DMR-1420570, ECS-0335765, ECCS-1542081).

**Acknowledgments**

We thank G. Zhong for generous help in using CNS facilities; and J. Treichler, A. Windsor, G. Bordonaro, K. Musa, and D. Botsch for their generous help in using CNF facilities. This work was performed in part at the Center for Nanoscale Systems (CNS), a member of the National Nanotechnology Infrastructure Network (NNIN), which is supported by the National Science Foundation under NSF award no. ECS-0335765. CNS is part of Harvard University. This work was performed in part at the Cornell NanoScale Facility (CNF), a member of the National Nanotechnology Coordinated Infrastructure (NNCI), which is supported by the National Science Foundation (Grant ECCS-1542081).